\RequirePackage{lineno}
\documentclass[twocolumn,letterpaper,aps,prc,longbibliography,superscriptaddress,nofootinbib,floatfix]{revtex4-1}

\usepackage{graphicx}	% Include figure files
\usepackage{appendix}
\usepackage{float}
\usepackage{bbm}
\usepackage{hyperref}
\usepackage{amsmath}
\usepackage{verbatim}
\usepackage{caption}

\hypersetup{
    colorlinks=true,
    linkcolor=blue,
    filecolor=blue,
    urlcolor=blue,
    citecolor=blue,
}

\usepackage{xspace}	% Include xspace

\newcommand{\pt}{\mbox{$p_T$}\xspace}

\newcommand{\pp}{\mbox{$p$$+$$p$}\xspace}
\newcommand{\dau}{\mbox{$d$$+$Au}\xspace}

\newcommand{\pau}{\mbox{$p$$+$Au}\xspace}

\newcommand{\pPb}{\mbox{$p$$+$Pb}\xspace}
\newcommand{\ppb}{\mbox{$p$$+$Pb}\xspace}
\newcommand{\heauthree}{\mbox{$^{3}$He$+$Au}\xspace}

\newcommand{\meanpt}{$\left< p_{\textrm{T}} \right>$\xspace}

\newcommand{\sonic}{\mbox{{\sc sonic}}\xspace}
\newcommand{\supersonic}{\mbox{{\sc sonic}}\xspace}
\newcommand{\music}{\mbox{{\sc music}}\xspace}
\newcommand{\ipglasma}{\mbox{{\sc ip-glasma}}\xspace}
\newcommand{\urqmd}{\mbox{{\sc urqmd}}\xspace}
\newcommand{\adscft}{\mbox{{\sc a}d{\sc s/cft}}\xspace}
\newcommand{\pip}{\mbox{$\Pi/\mathrm{P_{0}}$}\xspace}

\begin{document}

\title{The Skinny on Bulk Viscosity and Cavitation in Heavy Ion Collisions}

\newcommand{\colorado}{University of Colorado, Boulder, Colorado 80309, USA}
\affiliation{\colorado}

\newcommand{\pusan}{Pusan National University, Pusan, South Korea}

\author{M.~Byres} \affiliation{\colorado}
\author{S.H.~Lim} \affiliation{\pusan}
\author{C.~McGinn} \affiliation{\colorado}
\author{J. Ouellette} \affiliation{\colorado}
\author{J.L.~Nagle} \affiliation{\colorado}

\date{\today}

\begin{abstract}
Relativistic heavy ion collisions generate nuclear-sized droplets of quark-gluon plasma (QGP) that
exhibit nearly inviscid hydrodynamic expansion.    Smaller collision systems such as \pau, \dau, and \heauthree at 
the Relativistic Heavy Ion Collider, as well as \ppb and high-multiplicity \pp at the Large Hadron Collider 
may create even smaller droplets of QGP.   If so, the standard time evolution paradigm of heavy ion collisions
may be extended to these smaller systems.    These small systems present a unique opportunity to examine
pre-hydrodynamic physics and extract properties of the QGP, such as the bulk viscosity, where the short lifetimes
of the small droplets makes them more sensitive to these contributions.   Here we focus on the influence of bulk
viscosity, its temperature dependence, and cavitation effects on the dynamics in small and large systems using the publicly available
hydrodynamic codes \sonic and \music.   We also compare pre-hydrodynamic physics in different frameworks including
\adscft strong coupling, \ipglasma weak coupling, and free streaming or no coupling.
\end{abstract}

\pacs{25.75.Dw}

% For heavy ion papers we usually use just the one above (max is 4)
%%%%%%%%% Examples for p+p and spin papers include:
% PPG031:  \pacs{14.20.Dh, 13.60.Hb, 21.10.Hw, 25.40.Fq}
% PPG050:  \pacs{14.20.Dh, 25.40.Ep, 13.85.Ni, 13.88.+e}
% PPG037:  \pacs{13.85.Qk, 13.20.Fc, 13.20.He, 25.75.Dw}

\maketitle

%%%%%%%%%%%%%%%%%%%%%%%%%%%%%%%%%%%%%%%%%%%%%%%%%%%%%%%%%%%%%%%%%%%%%%%%%%%

%%%%%%%%%%%%%%%%%%%%%%%%%%%%%%%%%%%%%%%%%%%%%%%%%%%%%%%%%%%%%%%%%%%%%%%%%%%

\section{Introduction}
\label{sec:intro}

At temperatures exceeding $T > 150$~MeV, hadronic matter transitions to a state of deconfined quarks and gluons referred to as quark-gluon plasma (QGP). 
Some properties of QGP can be calculated via lattice QCD, for example the Equation of State (EOS) -- for a recent review see Ref.~\cite{Ratti:2018ksb}. 
In contrast other key properties such as the shear viscosity $\eta$ and bulk viscosity $\zeta$, typically normalized by the entropy density $s$, remain beyond current first principles calculations. 
There has been enormous attention paid to studies of the QGP shear viscosity to entropy density ratio $\eta/s$ in part because there is a conjectured lower bound of $1/4\pi$ derived within the context of string theoretical \adscft~\cite{Kovtun:2004de}. 
Extensive experimental measurements at the Relativistic Heavy Ion Collider (RHIC) and the Large Hadron Collider (LHC) indicate that the QGP value near the crossover temperature region $150 < T < 400$~MeV is close to the conjectured bound -- for a useful review see Ref.~\cite{Romatschke:2017ejr,Heinz:2013th}. 
In contrast, less attention has been cast on studies and experimental constraints of bulk viscosity, though it is no less fundamental a property of QGP.

Modern fluid dynamics is defined as an effective field theory for long-wavelength degrees of freedom. For an uncharged relativistic fluid, these are the energy density $\epsilon$ and fluid velocity $u^\mu$. Expectation values of operators such as the energy-momentum tensor $\langle T^{\mu\nu}\rangle$ in the underlying quantum field theory are expanded in powers of gradients of $\epsilon,u^\mu$, e.g.uncharged relativistic fluid, these are the energy density
\begin{equation}
\langle T^{\mu\nu}\rangle=T_{(0)}^{\mu\nu}+T_{(1)}^{\mu\nu}+\ldots\,,
\end{equation}
where $T^{\mu\nu}_{(0)}$ contains no gradients, $T^{\mu\nu}_{(1)}$ contains first-order gradients, and so on. One can show that for an uncharged fluid the first-order term contains only two different components,
\begin{equation}
\label{eq:tmunu1}
T^{\mu\nu}_{(1)}=\eta \sigma^{\mu\nu}+\zeta \left(\nabla_\lambda u^\lambda\right) \Delta^{\mu\nu}\,,
\end{equation}
where $\sigma^{\mu\nu}$ is the traceless shear-stress tensor that is first order in gradients, and $\Delta^{\mu\nu}$ is a symmetric projector that obeys $u_\mu \Delta^{\mu\nu}=0$ and $\Delta^\mu_\mu=3$.  
The form of (\ref{eq:tmunu1}) implies that the shear viscosity coefficient $\eta$ couples to shear stresses, while bulk viscosity $\zeta$ couples to expansion or compression of the fluid.

Conservation of the energy-momentum tensor (which is exact) leads to the equations of motion for the fields $\epsilon,u^\mu$, defining the theory. To lowest order in the gradient expansion (ideal fluid dynamics), one finds for instance 
\begin{equation}
u^\mu \nabla_\mu \epsilon+(\epsilon+P) \left(\nabla_\lambda u^\lambda\right)=0
%\frac{\eta}{2}\sigma_{\mu\nu} \sigma^{\mu\nu}+\zeta \left(\nabla_\lambda u^\lambda\right)^2\,.
\end{equation}
Many non-relativistic fluids are well approximated by assuming incompressibility, e.g. $\epsilon={\rm const.}$, which from the above equation of motion therefore implies $\nabla_\lambda u^\lambda=0$, hence the bulk-viscous term in (\ref{eq:tmunu1}) is absent even if $\zeta\neq 0$. Therefore, bulk viscosity does not play an important role in non-relativistic fluid dynamics.

On the other hand, relativistic fluids are never incompressible since Lorentz contractions must be exactly preserved. Therefore, bulk viscous effects can potentially be of importance for relativistic systems.

Since ${\rm Tr}\,T^{\mu\nu}_{(1)}=3\zeta \left(\nabla_\lambda u^\lambda\right)$, the bulk viscosity coefficient vanishes if the energy-momentum tensor is traceless, ${\rm Tr}\,\langle T^{\mu\nu}\rangle=0$. This is the case for systems with conformal symmetry, such that for conformal fluids $\zeta=0$. For non-conformal systems, $\zeta\neq 0$, and the value of $\zeta$ is related to the two-point correlation function of the trace of the energy-momentum tensor.

\section{Bulk Viscosity and Cavitation}

There are numerous works that have attempted to understand the bulk viscosity as a function of temperature motivated by various theoretical calculations -- for a comprehensive discussion see Refs.~\cite{Arnold:2006fz,Buchel:2007mf,Moore:2008ws,Lu:2011df,Dobado:2011qu,Meyer:2011gj,Czajka:2018bod}.    
General arguments follow the reasoning that the QGP becomes more conformal at very high temperature, as indicated by both perturbative QCD and lattice QCD calculations of the trace anomaly. This leads to the expectation that the bulk viscosity to entropy density goes to zero at high temperature, though this gives little guidance as to how quickly. Conversely, deep in the hadronic phase (T $<<$ 150 MeV), bulk viscosity is expected to become exponentially large, $\zeta\propto \frac{f_\pi^8}{m_\pi^5} e^{2 m_\pi/T}$ with $m_\pi$ and $f_\pi$ are the pion mass and decay constant, respectively \cite{Lu:2011df}. However, note that this result for the bulk viscosity depends on the time-scale one is interested in. In equilibrium (extremely long times), bulk viscosity is dominated by the rate of number changing processes, which are very slow (hence the large bulk viscosity coefficient). If one is interested in slightly out-of-equilibrium situations, the effective bulk viscosity coefficient is smaller \cite{Dobado:2011qu}. In a real-world heavy-ion collision, the relevant time-scale is much shorter, so number changing processes are not sufficiently fast to equilibrate particle species. As a result, the effective bulk viscosity in this low-temperature regime is extremely small (negligible). However, since particle species can no longer be exchanged, the system has frozen out chemically~\cite{Lu:2011df}, such that in lieu of a bulk viscosity there is a chemical potential for each frozen-out species. In Ref.~\cite{Torrieri:2007fb}, it has been suggested that bulk viscosity may have some relation with hadronization in heavy-ion collisions.

Using a phenomenological model to relate lattice gauge theory results for ${\rm Tr}\,\langle T^{\mu\nu}\rangle$ in QCD to the bulk viscosity coefficient, Ref.~\cite{Karsch:2007jc} reported a decline by an order of magnitude for $\zeta/s$ from $T_c$ to $1.2 T_c$.
Unfortunately, some of the bulk viscosity values reported in Ref.~\cite{Karsch:2007jc} were \textit{negative}, which would have indicated a thermodynamic instability of QCD.
In Ref.~\cite{Romatschke:2009ng}, it was later shown that the ad-hoc model used in Ref.~\cite{Karsch:2007jc} was based on an incorrect sum-rule.
As a result, there currently is no reliable constraint on the temperature dependence in the range accessible by present-day colliders.  
\begin{figure}
    \centering
    \includegraphics[width=\linewidth]{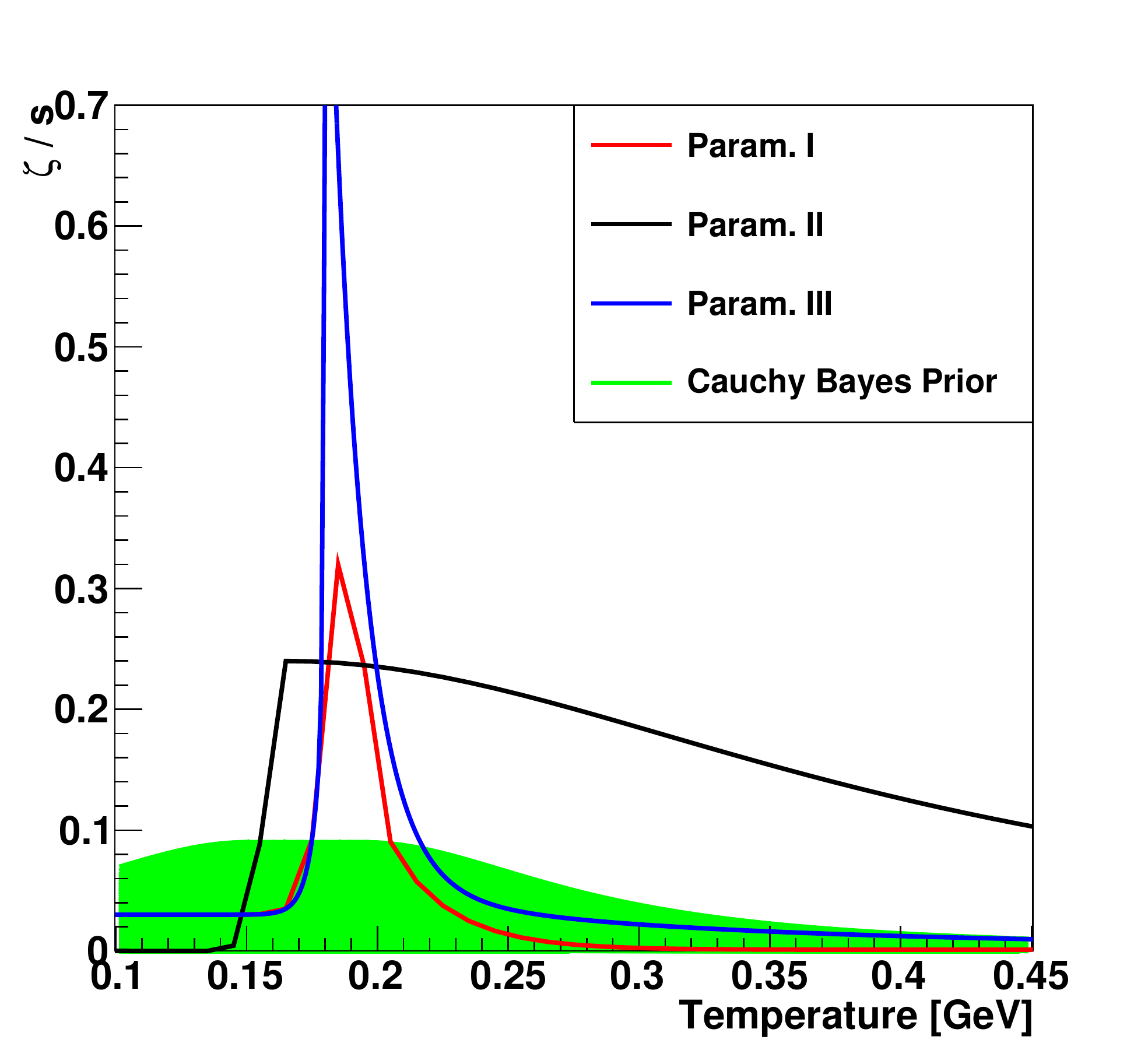}
    \caption{Various parameterizations for the temperature dependence of $\zeta/s$.   The green band represents the family of parameters of the Cauchy function utilized at the Bayesian Prior as detailed in Ref.~\cite{Moreland:2018gsh}.}
    \label{fig:bulk}
\end{figure}

Phenomenological studies at temperatures below the QCD transition, i.e. in the hadron gas regime including multiple resonance states, yield a bulk viscosity with a rather flat temperature dependence.  Inclusion of an exponential continuum of resonance states (a so-called ``Hagedorn State'' (HS) continuum~\cite{Hagedorn:1965st}) yields $\zeta/s$ values that modestly increase with temperature~\cite{NoronhaHostler:2008ju}.  However, this work~\cite{NoronhaHostler:2008ju} again uses the incorrect sum-rule model proposed in Ref.~\cite{Karsch:2007jc}.
Regardless, using results from Refs.~\cite{Karsch:2007jc,NoronhaHostler:2008ju} as input, the authors of Ref.~\cite{PhysRevC.97.034910} in turn introduced the following parameterization for $\zeta/s$ for use in hydrodynamic calculations:

\begin{equation}
\zeta/s=
\begin{cases} 
\begin{aligned} 0.03 + 0.9 e^{\frac{T/T_{p}-1}{0.0025}} + \\ 0.22e^{\frac{T/T_{p}-1}{0.022}} \end{aligned}  & T < 180~\textrm{MeV} \\ \\
\begin{aligned} 27.55 (T/T_{p}) - 13.45 \\ -13.77 (T/T_{p})^{2} \end{aligned} & 180 < T < 200 \\ \\
\begin{aligned} 0.001 + 0.9 e^{\frac{-(T/T_{p}-1)}{0.0025}} + \\
    0.25 e^{\frac{-(T/T_{p}-1)}{0.13}} \end{aligned} & T > 200~\textrm{MeV}
\end{cases}
\end{equation}
where $T_{p} = 180$~MeV.   This parameterization, labeled as ``Param. I'' peaks with a maximum $\zeta/s \approx 0.3$ and is shown in Figure~\ref{fig:bulk}.

Recently, an alternative parameterization has been put forth~\cite{Schenke:2019ruo} that has a large value for $\zeta/s$ over a very broad temperature range, also shown in Figure~\ref{fig:bulk} and labeled as ``Param. II''.

\begin{equation}
\zeta/s=
\begin{cases}
  B_{norm} {{B_{width}^{2}} \over {(T/T_{peak}-1)^{2} + B_{width}^{2}}} & T < T_{peak} \\ \\
  B_{norm} exp{ \left( -{{T-T_{peak}} \over {T_{width}}}^{2} \right) } & T > T_{peak}
\end{cases}
\end{equation}
The parameters are set with $T_{peak} = 165$~MeV, $B_{norm}$=0.24, $B_{width}=$1.5, and $T_{width}$= 10~MeV, where we note an error in Ref.~\cite{Schenke:2019ruo} misquoted $T_{width}$=50~MeV.
This new parameterization is determined within their hybrid (\ipglasma + hydrodynamic \music + hadronic scattering \urqmd) calculation primarily by the mean transverse momentum of midrapidity hadrons $<p_{T}>$ in heavy-ion $A+A$ collisions, highlighting the significant constraint from peripheral $A+A$ collisions. The \ipglasma~\cite{Schenke:2012wb} early stage of evolution follows weakly-coupled Yang-Mills dynamics and as such is close to free streaming.   Thus, large radial flow quickly develops, i.e. where the radial position and radial velocities are highly correlated.   This is particularly true in smaller collision volumes where the length scale of pre-hydrodynamic expansion $c \tau$ is on the same scale as the overall size of the system.   Thus, a larger bulk viscosity is needed to temper the large radial flow that the hydrodynamics is initialized with -- as noted in Ref.~\cite{Schenke:2018fci}.   

\begin{figure*}[ht]
    \centering
    \includegraphics[width=0.99\linewidth]{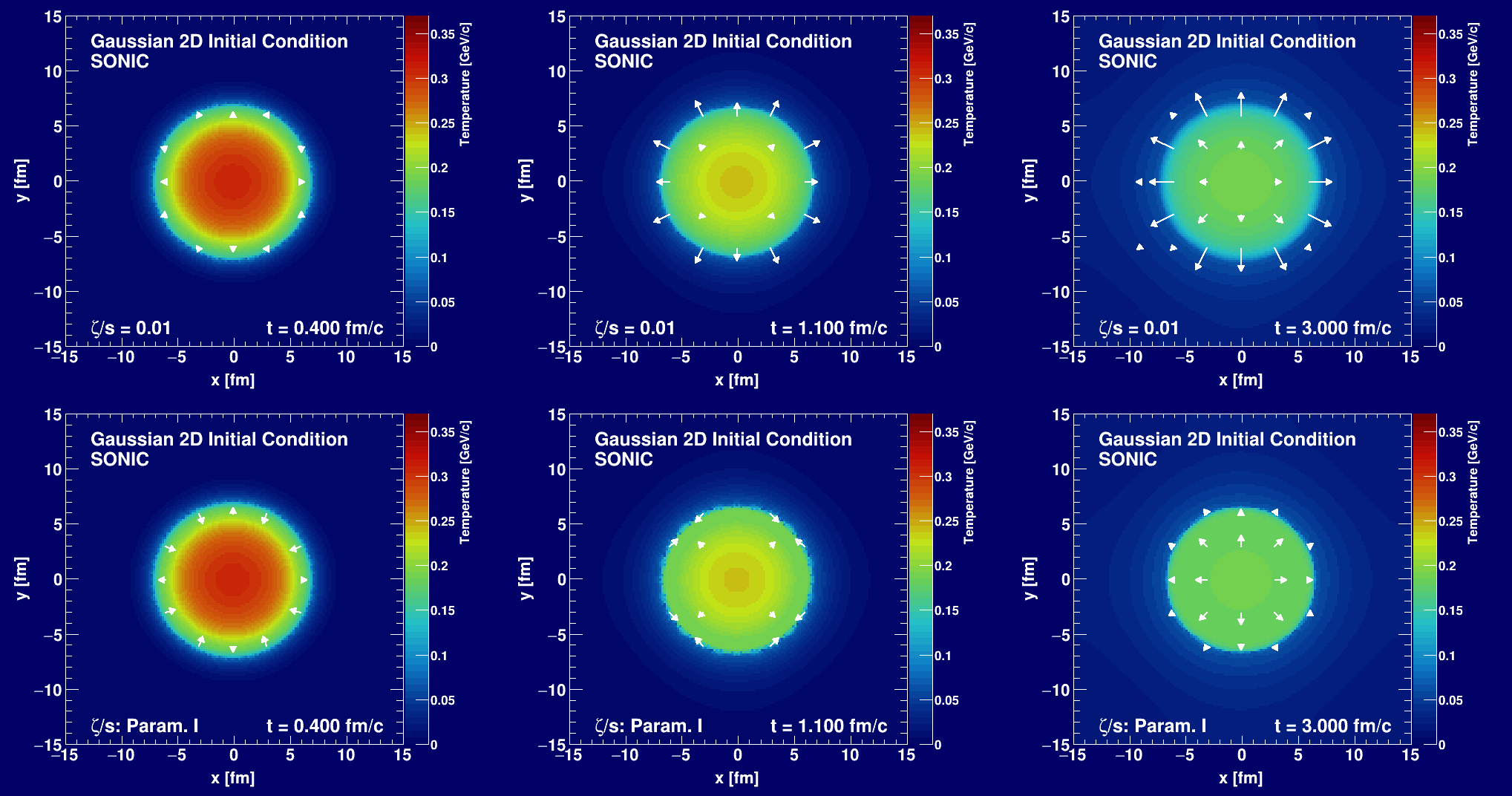}
    \caption{\sonic hydrodynamic simulation with single Gaussian source and $\eta/s = 1/4\pi$ and $\zeta/s=0.01$ (left column) or $\zeta/s$ from Param. I (right column).   Arrows represent the fluid cell velocities.}
    \label{fig:snapshot1}
\end{figure*}

In peripheral $A+A$ collisions, experimental data indicates a decrease in the \meanpt, whereas without the large bulk viscosity the hybrid calculation would have a modest increase.   We highlight that using peripheral $A+A$ data for \meanpt is likely incorrect as recent studies indicate that a centrality selection bias works to anti-select against jets in these events~\cite{Morsch:2017brb}.   Jets increase the multiplicity and move events to slightly more central categories, and thus more peripheral categories see a suppression of higher $p_T$ particles~\cite{Acharya:2018njl}.   We also note that in this hybrid approach, the transition temperature between hydrodynamics and hadron cascade is set at 145~MeV, corresponding to a point where $\zeta/s$ is actually quite small.  
This parameterization has also been applied to hydrodynamic calculations for small collision systems~\cite{Schenke:2019pmk} -- which is very relevant to this paper as the interplay of pre-hydrodynamic physics and QGP properties becomes critical.

We highlight two other parameterizations considered in the literature for comparison.   There is an additional parameterization similar to Param. I in that it is sharply peaked near the transition temperature.   This parameterization, referred to as ``Param. III'' is shown in Figure~\ref{fig:bulk} and peaks with an even larger value of $\zeta/s$~\cite{Denicol:2015bpa}.

Lastly, in a Bayesian inference analysis of \pPb and Pb+Pb flow patterns~\cite{Moreland:2018gsh},
the authors include a parameterization of $\zeta/s$ using a Cauchy function:
\begin{equation}
    (\zeta/s)(T) = {{(\zeta/s)_{max}} \over {1 + ( (T-(\zeta/s)_{T_{0}}) / (\zeta/s)_{width})^{2}}}. 
\end{equation}
In their Bayesian prior, they consider these parameters over the following ranges:
$(\zeta/s)_{max}$ = 0.0 -- 0.1, $(\zeta/s)_{T_{0}}$ = 150 -- 200~MeV, and 
$(\zeta/s)_{width}$ = 0 -- 100~MeV.   The full range of functional forms with these
parameter ranges is shown as a green band in Figure~\ref{fig:bulk}.    It is striking that
the Bayesian prior is rather restrictive and completely excludes the three other
parameterizations used elsewhere in the current literature.

Both shear and bulk viscosity represent out-of-equilibrium corrections to ideal hydrodynamics.  At the freeze-out point in hydrodynamics, the standard approach is to transition to particles via the Cooper-Frye formalism~\cite{Cooper:1974mv} which explicitly utilizes the equilibrium condition.  In the case of large shear or bulk viscosity, the hadronization formalism requires non-equilibrium corrections, which are parametrized by a $\delta f$ term. For small departures from equilibrium, the $\delta f$ term may be obtained by matching to kinetic theory, cf. Refs.~\cite{Teaney:2003kp,Romatschke:2017ejr}
\begin{equation}
\delta f \propto f_{\rm eq}\frac{p^\mu p^\nu}{T p^\rho u_\rho}\left(\frac{\sigma_{\mu\nu}}{2}+\left(\frac{\Delta_{\mu\nu}}{3}-c_s^2 u_\mu u_\nu\right) \nabla_\lambda u^\lambda\right)\,.
\end{equation}
   
For conformal systems where the bulk correction vanishes, a model for $\delta f$ can be derived that correctly reproduces the hydrodynamic energy-momentum tensor and is well behaved even for large shear stresses (cf. ~\cite{Romatschke:2017ejr}, section 3.1.5). Similarly, models for $\delta f$ within the framework of anisotropic hydrodynamics have been proposed that correctly reproduce a non-interacting expanding gas \cite{Strickland:2014pga}.

However, in the case of bulk viscosity, the $\delta f$ correction is not well known \cite{Molnar:2014fva}. Thus, in some works with significant non-zero $\zeta/s$, they simply apply no $\delta f$ correction at all. Ref.~\cite{Bernhard:2016tnd} addresses this as follows: "Given this uncertainty and the small
$\zeta/s$ at particlization, we assume that bulk
corrections will be small and neglect them for the present study, i.e. $\delta f$ bulk = 0. This precludes any quantitative conclusions on bulk viscosity, since we are only allowing bulk viscosity to affect the hydrodynamic evolution, not particlization."    A major issue here is that there should be a systematic uncertainty in the calculations for the temperature at which one transitions from hydrodynamics to hadron cascade.   Of course, by not varying this temperature, one only considers cases where hadronization occurs after the $\zeta/s$ value has fallen by more than a factor of ten from its peak at the nominal transition temperature.    As we explore in this paper, this leads to enormous uncertainties, particularly for small collision systems.

In contrast, other calculations utilize specific examples of possible bulk $\delta f$ parameterizations -- see for example Ref.~\cite{PhysRevC.97.034910}.   In general they also only consider variations in hadronization temperatures in the lower temperature region when $\zeta/s$ is very small.   In the \supersonic calculations~\cite{Weller:2017tsr}, they only consider very small values of $\zeta/s=0.01$ (discussed more below) and thus neglect any
$\delta f$ correction.   We highlight here that they implement a strongly-coupled \adscft inspired pre-hydrodynamic evolution.    As such, in contrast to free streaming or \ipglasma evolution, no extreme radial flow develops early and hence no large bulk viscosity is required to temper the growth in $<p_{T}>$, as was explicitly shown Figure 5 of Ref.~\cite{vanderSchee:2013pia}.

A critical issue with implementations of large bulk viscosity relate to cavitation.  Cavitation is the formation of bubbles or cavities within a fluid, appearing in areas of low relative pressure. In case of QGP droplets, cavitation implies regions of hadronic gas inside the fluid at temperatures well above T$_{\mathrm{c}}$. These bubbles can grow, shrink, collide and otherwise interacted in a highly non-trivial manner, and the physics of cavitation is not described in current hydrodynamic codes -- for example \music, \sonic, or any other heavy-ion implementation.

Cavitation in heavy-ion collisions was first discussed in detail in Ref.~\cite{Rajagopal:2009yw}.   The exploratory study concluded that $\zeta/s$ with a significant enough peak near the transition temperature will trigger cavitation.
A \supersonic calculation detailed in Ref.~\cite{Habich:2015rtj} calculates LHC energy $p+p$ collisions and a bulk viscosity $\zeta/s \approx 0.01-0.02$ is required to moderate the growth in $<p_{T}>$ in high-multiplicity events.  These small values of $\zeta/s$ yield small values of the bulk pressure $\Pi$ and thus cavitation does not occur.

\begin{figure*}
    \centering
    \includegraphics[width=0.95\linewidth]{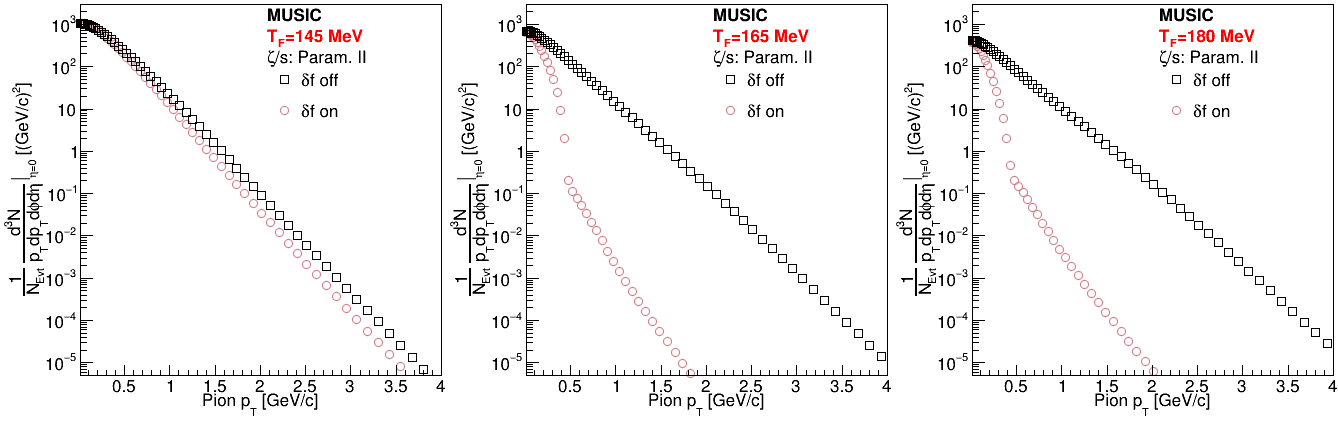}
    \caption{\music results for a single Gaussian initial geometry run with $\zeta/s$ Param. II with three different freeze-out temperatures.   Shown are pion \pt distributions with and without $\delta f$ corrections applied.}
    \label{fig:ptexample}
\end{figure*}

\begin{table*}[]
 \centering
 \begin{tabular}{|c|c|c|c|c|c|} \hline
  Test case & Bulk Param. & $\delta f$ corr. & $T_{hadronize}$ & $<p_{\mathrm{T}}>$ pion & $<p_{\mathrm{T}}>$ proton  \\ \hline \hline
  0 & Param. I & yes & 145 MeV & 413~MeV & 790~MeV \\ \hline
  1 & Param. I & no & 145 MeV & 494~MeV & 853~MeV \\ \hline
  2 & Param. I & yes & 165 MeV & 382~MeV & 698~MeV \\ \hline
  3 & Param. I & no & 165 MeV & 496~MeV & 810~MeV \\ \hline
  4 & Param. I & yes & 180 MeV & 145~MeV & 617~MeV \\ \hline
  5 & Param. I & no & 180 MeV & 545~MeV & 889~MeV \\ \hline
  6 & Param. II & yes & 145 MeV & 406~MeV & 706~MeV \\ \hline
  7 & Param. II & no & 145 MeV & 450~MeV & 735~MeV \\ \hline
  8 & Param. II & yes & 165 MeV & 170~MeV & 534~MeV \\ \hline
  9 & Param. II & no & 165 MeV & 466~MeV & 742~MeV \\ \hline
  10 & Param. II & yes & 180 MeV & 151~MeV & 532~MeV \\ \hline
  11 & Param. II & no & 180 MeV & 499~MeV & 788~MeV \\ \hline
 \end{tabular}
 \caption{\music results for the single Gaussian geometry for the pion and proton \meanpt for different $\zeta/s$ parameterizations, turning on and off $\delta f$ corrections, and different freeze-out temperatures.}
 \label{tab:numbers}
\end{table*}

In contrast, in Ref.~\cite{Denicol:2015bpa}, the authors consider large $\zeta/s$ scenarios corresponding to Param. I and Param. III, and find that for $A+A$ central collisions, Param.
I leads to modest regions of cavitation at large radius at early times and in Param. III 
leads to significant cavitation over a very large portion of the space-time volume of QGP.  
They plot the ratio of bulk pressure $\Pi$ divided by the standard pressure 
$\mathrm{P_{0}}$, \pip. If \pip falls below 1.0, then the effective pressure is negative and cavitation is possible.

In the calculations with Param. II that include large bulk viscosities over a wide range of temperatures, the issue of cavitation is not discussed~\cite{Schenke:2019ruo, Schenke:2019pmk}.   A focus of this paper is to explore the issue of cavitation in these various scenarios.

Noting that large $\zeta/s$ is likely to lead to unreliable hydrodynamic results as well as very large systematic uncertainties from variations in the hadronization temperature and $\delta f$ corrections, that does not logically lead to the conclusion that these large $\zeta/s$ values are incorrect.   Mother nature does not have to be kind.   

However, if cavitation does indeed correspond to the onset of hadronization as advocated for in Ref.~\cite{Romatschke:2017ejr}, then its occurrence should be observable.  According to the argument outlined in Ref.~\cite{Habich:2014tpa}, cavitation would lead to the creation of hadrons at temperatures well above the QCD phase transition, which would be detected. Since experimentally measured hadron spectra are inconsistent with freeze-out temperatures much in excess of $T_c$, this implies that significant cavitation does not occur in heavy-ion collisions. As a consequence, bulk viscosity values that lead to strong cavitation events in hydrodynamic simulations of ion collisions are likely disallowed by experiment.
%%%%%%%
%%%%%%%%%%%%%%%%%%%%%%%%%%%%%%%%%%%%%%%%%%%%%%%%%%%%%%%%%%%%%%%%%%%%%%%%%%%

\section{Definition of Methods}
\label{sec:method}

In this study, we are not interested in exact matching of experimental data and rather understanding the basic consequences of different bulk viscosity implementations.    To that end, we consider a simple geometry in our studies.   
We utilize a geometry defined by
two-dimensional Gaussian distribution for the energy density
\begin{equation}
    \epsilon (x,y) = \epsilon_{0} e^{-(x^{2}+y^{2})/2\sigma^{2}}
\end{equation}
where the width is set $\sigma = 1.0$~fm and the normalization $\epsilon_{0}$ is set such that the corresponding central temperature $T \approx 370$~MeV.  This initial energy density is a proxy for small collision systems, and one where we have chosen an azimuthally symmetric parameterization to focus on radial expansion.

\begin{figure*}[ht]
    \centering
    \includegraphics[width=0.95\linewidth]{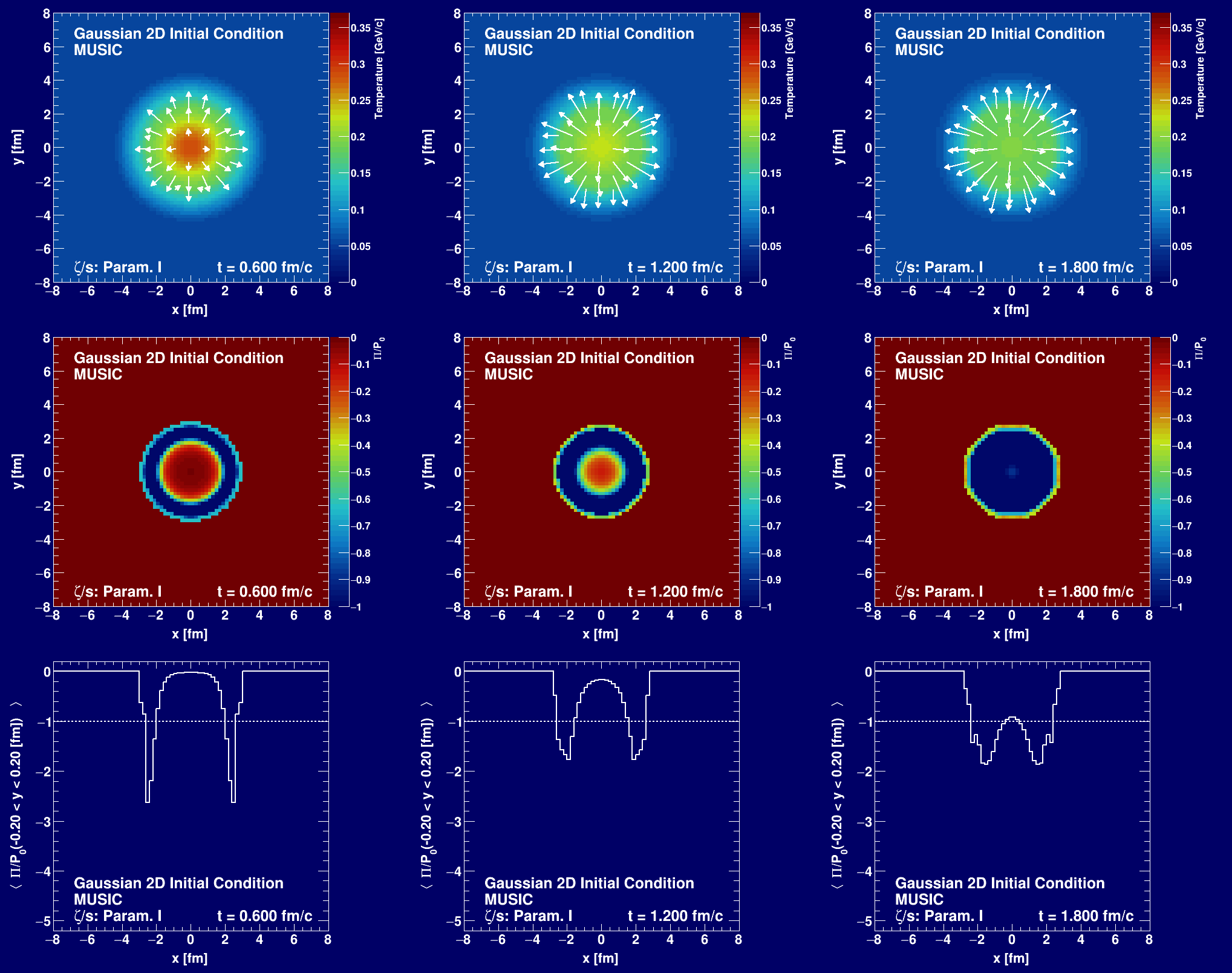}
    \caption{\music hydrodynamic simulation with single Gaussian source and $\eta/s = 1/4\pi$ and $\zeta/s$ as Param. I. Left to right is a time progression from 0.6 to 1.2 to 1.8 fm/c, while top panels are temperature profile, middle panels are \pip, and bottom panels are a 1-D slice of \pip across middle region -0.2 $<$ y $<$ 0.2 fm/c. In middle panels, darkest blue cells indicate regions of cavitation, also indicated in bottom panels by values of \pip $<$ -1.}
    \label{fig:musicParamITimeProg}
\end{figure*}

We have utilized two publicly available hydrodynamic codes, namely \sonic~\cite{Romatschke:2017ejr} and \music~\cite{Schenke:2010nt}.    Both codes provide numerical solutions to relativistic viscous hydrodynamics and include as inputs temperature-dependent parameterizations for the shear $\eta/s$ and bulk $\zeta/s$ to entropy density ratios.   In the case of \music, we have run the code in 2+1 dimension mode, i.e. not invoking the 3+1 dimension option.   Also, \music has a $\delta f$ correction option that can be turned on and off for shear and bulk viscosity.  In contrast there is no bulk $\delta f$ correction option in \sonic.   In both cases, we are simply running the hydrodynamics starting at $\tau_{0}=0.2$~fm/c with the initial energy density $\epsilon (x,y)$, i.e. no pre-hydrodynamic evolution, and then with no post-hydrodynamic hadronic re-scattering.

\section{Results}

Utilizing the \sonic code, time snapshots of the two-dimensional temperature profile from \sonic for the simple Gaussian geometry run with $\eta/s=1/4\pi$ (temperature independent) and $\zeta/s=0.01$ (also temperature independent) are shown in Figure~\ref{fig:snapshot1} (left column).   The arrows represent a sampling of the fluid cell velocities.   A significant radial outward flow quickly develops before the system reaches the $T=145$~MeV user-defined freeze-out temperature.
In contrast, shown in Figure~\ref{fig:snapshot1} (right column) for the same times are \sonic results with the same simple Gaussian initial geometry with a significant bulk viscosity given by Param. I, with peak  $\zeta/s \approx 0.3$.    The result is a substantial reduction in the transverse expansion as much of the system simply cools via longitudinal expansion.    
The velocity arrows are significantly smaller and the overall radial extent of the system is correspondingly smaller at later times.

\begin{figure*}[ht]
    \centering
    \includegraphics[width=0.95\linewidth]{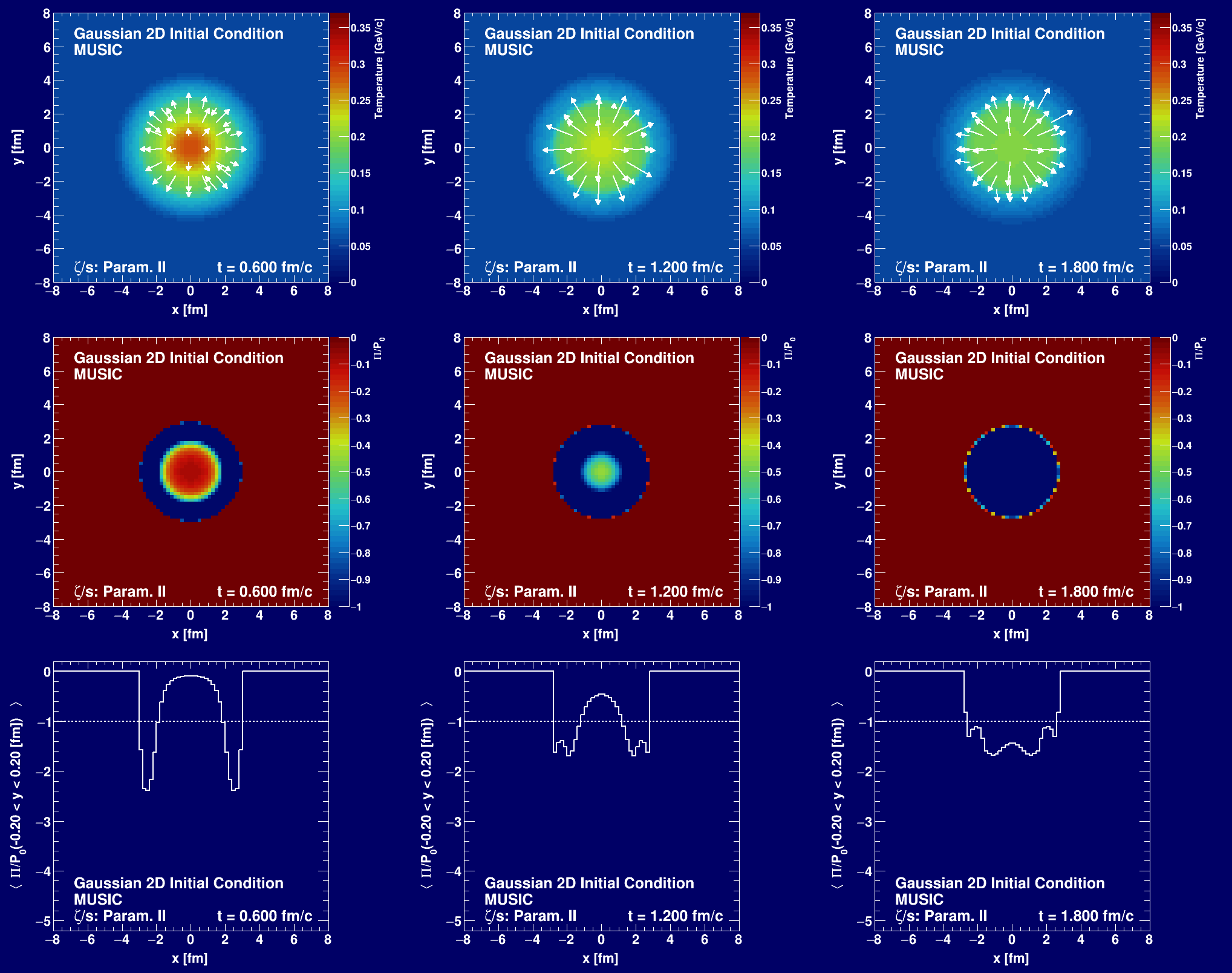}
    \caption{\music hydrodynamic simulation with single Gaussian source and $\eta/s = 1/4\pi$ and $\zeta/s$ as Param. II. Left to right is a time progression from 0.6 to 1.2 to 1.8 fm/c, while top panels are temperature profile, middle panels are \pip, and bottom panels are a 1-D slice of \pip across middle region -0.2 $<$ y $<$ 0.2 fm/c. In middle panels, darkest blue cells indicate regions of cavitation, also indicated in bottom panels by values of \pip $<$ -1.}
    \label{fig:musicParamIITimeProg}
\end{figure*}

In the case of large $\zeta/s$ the mean transverse momentum \meanpt for resulting hadrons is significantly reduced, as discussed in detail later.    Thus the bulk viscosity has a substantial effect on the overall space-time evolution as well as the resulting distribution of hadrons.
In contrast, \sonic studies with \adscft pre-hydrodynamic evolution for p+p collision geometries, a modest $\zeta/s = 0.01$ (implemented as temperature independent) was found to slightly reduce the \meanpt from initial radial flow in agreement with experimental data~\cite{Weller:2017tsr}.   

We have confirmed that running \sonic and \music with the same initial geometry, the same $\zeta/s$ parameterization, and bulk $\delta f$ correction turned off (since \sonic does not have an implementation for bulk viscosity related $\delta f$) gives good agreement for the pion and proton \pt distributions.   Hence, using \music we can further explore the dependencies of the different $\zeta/s$ parameterizations in conjunction with turning on and off the $\delta f$ correction.   
Shown in Figure~\ref{fig:ptexample} are the pion \pt distributions at three different freeze-out temperatures ($T_{f} = 145, 165, 180$~MeV from left to right) for 
$\zeta/s$ parameterization Param. II with and without $\delta f$ corrections.   The change in the 
pion \pt with and without $\delta f$ corrections is modest when $T_{f} = 145$~MeV since as seen in Figure~\ref{fig:bulk} the value at that temperature for $\zeta/s < 0.01$.    However, as soon as one changes $T_{f} = 160$~MeV, there is an enormous change in the \pt distribution since the bulk viscosity is already very large.   Table~\ref{tab:numbers} includes a full summary of pion and proton mean \pt values corresponding to $\zeta/s$ Param. I and II, three different freeze-out temperatures, and with and without $\delta f$ corrections.   The systematic uncertainty based on bulk-viscous non-equilibrium corrections via $\delta f$ are very large, often of order 100\% on the mean \pt, unless applied where the $\zeta/s$ parameterization yields very small values on the low temperature side (i.e. $T < 170$~MeV for Param. I and $T < 150$~MeV for Param. II).

These results highlight that if one exclusively restricts the freeze-out temperature to $T=145$~MeV, one can quote a rather small systematic uncertainty due to the lack of complete knowledge of the $\delta f$ correction, i.e. by comparing with and without its inclusion.    However, that is only because the parameterizations Param. I, II, and III all have rather small values of $\zeta/s$ at that temperature, specifically $\zeta/s=0.03, 0.0044, 0.03$, respectively.
It is reasonable to simultaneously systematically vary the freeze-out temperature, as that is physically allowed, and then from Table~\ref{tab:numbers} one observes enormous systematic variations in the \meanpt values. One cannot avoid these systematic uncertainties in a full evaluation.

Given these large values for $\zeta/s$ it is critical to assess whether the total effective pressure becomes negative and hence cavitation is unavoidable.  Figures~\ref{fig:musicParamITimeProg} and~\ref{fig:musicParamIITimeProg} show for three different time snapshots the temperature distributions along with fluid velocity vectors (upper row), the \pip distributions (middle row), and a one-dimensional slice along $y=0$ of the \pip values (lower row).
Figure~\ref{fig:musicParamITimeProg} corresponds to the sharply peaked $\zeta/s$ Param. I and Figure~\ref{fig:musicParamIITimeProg} to the wide $\zeta/s$ Param. II.    
We highlight that the same negative pressures are confirmed when running \sonic with $\zeta/s$ Param. I since the results are not specific to the numerical implementations in \music versus \sonic.

First focusing on the results as used in Refs.~\cite{Schenke:2019ruo,Schenke:2019pmk}, the vast majority of the space-time volume has \pip $< -1$ and hence the entire volume will undergo cavitation.   
The issue of cavitation is not mentioned in Refs.~\cite{Schenke:2019ruo,Schenke:2019pmk}, though we have confirmed with the authors that for their specific geometry (e.g. $p+A$) most of the space-time volume is in fact in the cavitation regime. 
This means that the results of the hydrodynamic evolution are  unreliable as they are well outside the equations domain of validity.    For the Param. I case, the cavitation effect is slightly less, but still placing the results outside the domain of hydrodynamics.

In an earlier check on cavitation in Pb+Pb collision geometries~\cite{Denicol:2015bpa}, the authors find that with the most sharply peaked Param. III, cavitation dominates the space-time volume.  We have confirmed these results with our simulations.    However, using Param. I they find a smaller space-time volume in the cavitation regime and then conclude that ``cavitation will not happen.''  Here caution is warranted since even if only a minority, e.g. 25\% of the space-time volume has \pip $< -1$, is in the cavitation regime, it is not possible to reliably conclude that the overall hydrodynamic results are robust.

\section{Discussion}

There are many recent cases where hydrodynamics appears to be applied outside its ``domain of validity.''   For example, there are often large $\delta f$ corrections even for shear viscosity.   However, the case of cavitation is worse because it is a known mechanical instability of the system, whereas large $\delta f$ merely implies that one did not use a good ``model'' for the transition from hydrodynamics to particles.  Cavitation says ''STOP: you are out of bounds.''   
Why does hydrodynamics not handle such regions of low pressure, i.e. cavities?   The reason is because these cavities are regions of a gas.   One can argue that one sometimes models the hadronic gas using hydrodynamics, and that is true, but here the error one is making is of order one.    A better approach would be to treat the cells that cavitate as ''frozen out'' with whatever their local temperature would be. For massive cavitation for cells with $T > 250$~MeV, the corresponding spectra will look unlike anything measured.   That may simply be indicating that your chosen bulk viscosity is inconsistent with experiment.

Recent developments have argued that hydrodynamics is valid over a much larger domain than previously assumed.   However, the argument is that one loses the hydrodynamic attractor if the system cavitates.   One can also break hydrodynamics with just shear stress, but the point is that cavitation is the reaction to a massive bulk stress, not a small perturbation.  The instability occurs whenever the pressure drops below the vapor pressure. Negative pressure is much more extreme, meaning that once the pressure is negative, the system is essentially guaranteed to have cavitation.

We recall again that mother nature does not have to be kind.    In the case of \sonic with \adscft pre-hydrodynamic evolution, i.e. strong coupling, there is no very strong build up of initial radial flow.    Hence, only a modest, temperature independent bulk viscosity $\zeta/s = 0.01$ is needed for matching small system \meanpt~\cite{Weller:2017tsr}.    In this case, there is almost no sensitivity to inclusion of a $\delta f$ correction or not, as well as little sensitivity to the transition temperature for ending hydrodynamics and starting the hadronic re-scattering.    

In contrast, the \ipglasma initial conditions and pre-hydrodynamic evolution appears to require a large bulk viscosity.   This raises some crisp questions or set of scenarios.    (1)  If such a large bulk viscosity can be ruled out as put forward in Ref.~\cite{Habich:2014jna}, then \ipglasma initial conditions would be ruled out.   (2)  If we consider large bulk viscosity as possible, then how can one arrive at reliable space-time evolution since the currently solved hydrodynamic equations do not include critical cavitation, (3) Is there a way to reconcile \ipglasma initial conditions without a large bulk viscosity?   

In order to evaluate these scenarios a set of apples-to-apples comparisons would be most fruitful.  This means having identical initial conditions and matching conditions with a code module interchange for \ipglasma, free streaming, and \adscft evolution prior to hydrodynamics.   That way the key features in the pre-hydrodynamic evolution necessitating a large bulk viscosity can be discerned.

%%%%%%%%%%%%%%%%%%%%%%%%%%%%%%%%%%%%%%%%%%%%%%%%%%%%%%%%%%%%%%%%%%%%%%%%%%%
\section{Conclusions}
\label{sec:conclusion}
In conclusion, we find utilizing the publicly available \sonic and \music hydrodynamic codes that currently used parameterizations of $\zeta/s$ result in large space-time volumes with negative effective pressure.    Thus, the systems will undergo cavitation, which is outside the domain of validity for hydrodynamics.     All implementations of $\zeta/s$ with different geometries need to test for cavitation.   Currently it is not possible to assess the full systematic uncertainty arising from ignoring these effects.   Future work to isolate the key initial conditions and pre-hydrodynamic evolution that necessitates in some models large bulk viscosity are critical to further progress.

%%%%%%%%%%%%%%%%%%%%%%%%%%%%%%%%%%%%%%%%%%%%%%%%%%%%%%%%%%%%%%%%%%%%%%%%%%%
\section*{Acknowledgments}
We acknowledge Paul Romatschke for semi-infinite useful discussions, support of the publicly available \sonic code, and a careful reading of the manuscript.   We acknowledge Bj$\ddot{\rm o}$rn Schenke and Chun Shen for useful discussions and support of the publicly available \music code.   We acknowledge Jean-Yves Ollitrault for a careful reading of the manuscript.  We acknowledge useful discussions on the subject with Dennis Perepelitsa and Bill Zajc.
MB, CM, JLN acknowledges support from the U.S. Department of Energy, Office of Science, Office of Nuclear Physics under Contract No. DE-FG02-00ER41152.   J.O. acknowledges support from the U.S. Department of Energy, Office of Science, Office of Nuclear Physics under Contract No. de-sc0018117.
SHL acknowledges support from a 2-Year Research Grant of Pusan National University.
%\clearpage
%%%%%%%%%%%%%%%%%%%%%%%%%%%  Appendix
\appendix

%%%%%%%%%%%%%%%%%%%%%%%%%%%

%%%%%%%%%%%%%%%%%%%%%%%%%%%  References
\clearpage

\bibliography{main}

%merlin.mbs apsrev4-1.bst 2010-07-25 4.21a (PWD, AO, DPC) hacked
%Control: key (0)
%Control: author (0) dotless jnrlst
%Control: editor formatted (1) identically to author
%Control: production of article title (0) allowed
%Control: page (1) range
%Control: year (0) verbatim
%Control: production of eprint (0) enabled
\begin{thebibliography}{37}%
\makeatletter
\providecommand \@ifxundefined [1]{%
 \@ifx{#1\undefined}
}%
\providecommand \@ifnum [1]{%
 \ifnum #1\expandafter \@firstoftwo
 \else \expandafter \@secondoftwo
 \fi
}%
\providecommand \@ifx [1]{%
 \ifx #1\expandafter \@firstoftwo
 \else \expandafter \@secondoftwo
 \fi
}%
\providecommand \natexlab [1]{#1}%
\providecommand \enquote  [1]{``#1''}%
\providecommand \bibnamefont  [1]{#1}%
\providecommand \bibfnamefont [1]{#1}%
\providecommand \citenamefont [1]{#1}%
\providecommand \href@noop [0]{\@secondoftwo}%
\providecommand \href [0]{\begingroup \@sanitize@url \@href}%
\providecommand \@href[1]{\@@startlink{#1}\@@href}%
\providecommand \@@href[1]{\endgroup#1\@@endlink}%
\providecommand \@sanitize@url [0]{\catcode `\\12\catcode `\$12\catcode
  `\&12\catcode `\#12\catcode `\^12\catcode `\_12\catcode `\%12\relax}%
\providecommand \@@startlink[1]{}%
\providecommand \@@endlink[0]{}%
\providecommand \url  [0]{\begingroup\@sanitize@url \@url }%
\providecommand \@url [1]{\endgroup\@href {#1}{\urlprefix }}%
\providecommand \urlprefix  [0]{URL }%
\providecommand \Eprint [0]{\href }%
\providecommand \doibase [0]{http://dx.doi.org/}%
\providecommand \selectlanguage [0]{\@gobble}%
\providecommand \bibinfo  [0]{\@secondoftwo}%
\providecommand \bibfield  [0]{\@secondoftwo}%
\providecommand \translation [1]{[#1]}%
\providecommand \BibitemOpen [0]{}%
\providecommand \bibitemStop [0]{}%
\providecommand \bibitemNoStop [0]{.\EOS\space}%
\providecommand \EOS [0]{\spacefactor3000\relax}%
\providecommand \BibitemShut  [1]{\csname bibitem#1\endcsname}%
\let\auto@bib@innerbib\@empty
%</preamble>
\bibitem [{\citenamefont {Ratti}(2018)}]{Ratti:2018ksb}%
  \BibitemOpen
  \bibfield  {author} {\bibinfo {author} {\bibfnamefont {Claudia}\ \bibnamefont
  {Ratti}},\ }\bibfield  {title} {\enquote {\bibinfo {title} {{Lattice QCD and
  heavy ion collisions: a review of recent progress}},}\ }\href {\doibase
  10.1088/1361-6633/aabb97} {\bibfield  {journal} {\bibinfo  {journal} {Rept.
  Prog. Phys.}\ }\textbf {\bibinfo {volume} {81}},\ \bibinfo {pages} {084301}
  (\bibinfo {year} {2018})},\ \Eprint {http://arxiv.org/abs/1804.07810}
  {arXiv:1804.07810 [hep-lat]} \BibitemShut {NoStop}%
%%CITATION = ARXIV:1804.07810;%%
\bibitem [{\citenamefont {Kovtun}\ \emph {et~al.}(2005)\citenamefont {Kovtun},
  \citenamefont {Son},\ and\ \citenamefont {Starinets}}]{Kovtun:2004de}%
  \BibitemOpen
  \bibfield  {author} {\bibinfo {author} {\bibfnamefont {P.}~\bibnamefont
  {Kovtun}}, \bibinfo {author} {\bibfnamefont {Dan~T.}\ \bibnamefont {Son}}, \
  and\ \bibinfo {author} {\bibfnamefont {Andrei~O.}\ \bibnamefont
  {Starinets}},\ }\bibfield  {title} {\enquote {\bibinfo {title} {{Viscosity in
  strongly interacting quantum field theories from black hole physics}},}\
  }\href {\doibase 10.1103/PhysRevLett.94.111601} {\bibfield  {journal}
  {\bibinfo  {journal} {Phys. Rev. Lett.}\ }\textbf {\bibinfo {volume} {94}},\
  \bibinfo {pages} {111601} (\bibinfo {year} {2005})},\ \Eprint
  {http://arxiv.org/abs/hep-th/0405231} {arXiv:hep-th/0405231 [hep-th]}
  \BibitemShut {NoStop}%
%%CITATION = HEP-TH/0405231;%%
\bibitem [{\citenamefont {Romatschke}\ and\ \citenamefont
  {Romatschke}(2019)}]{Romatschke:2017ejr}%
  \BibitemOpen
  \bibfield  {author} {\bibinfo {author} {\bibfnamefont {Paul}\ \bibnamefont
  {Romatschke}}\ and\ \bibinfo {author} {\bibfnamefont {Ulrike}\ \bibnamefont
  {Romatschke}},\ }\href {\doibase 10.1017/9781108651998} {\emph {\bibinfo
  {title} {{Relativistic Fluid Dynamics In and Out of Equilibrium}}}},\
  Cambridge Monographs on Mathematical Physics\ (\bibinfo  {publisher}
  {Cambridge University Press},\ \bibinfo {year} {2019})\ \Eprint
  {http://arxiv.org/abs/1712.05815} {arXiv:1712.05815 [nucl-th]} \BibitemShut
  {NoStop}%
%%CITATION = ARXIV:1712.05815;%%
\bibitem [{\citenamefont {Heinz}\ and\ \citenamefont
  {Snellings}(2013)}]{Heinz:2013th}%
  \BibitemOpen
  \bibfield  {author} {\bibinfo {author} {\bibfnamefont {Ulrich}\ \bibnamefont
  {Heinz}}\ and\ \bibinfo {author} {\bibfnamefont {Raimond}\ \bibnamefont
  {Snellings}},\ }\bibfield  {title} {\enquote {\bibinfo {title} {{Collective
  flow and viscosity in relativistic heavy-ion collisions}},}\ }\href {\doibase
  10.1146/annurev-nucl-102212-170540} {\bibfield  {journal} {\bibinfo
  {journal} {Ann. Rev. Nucl. Part. Sci.}\ }\textbf {\bibinfo {volume} {63}},\
  \bibinfo {pages} {123--151} (\bibinfo {year} {2013})},\ \Eprint
  {http://arxiv.org/abs/1301.2826} {arXiv:1301.2826 [nucl-th]} \BibitemShut
  {NoStop}%
%%CITATION = ARXIV:1301.2826;%%
\bibitem [{\citenamefont {Arnold}\ \emph {et~al.}(2006)\citenamefont {Arnold},
  \citenamefont {Dogan},\ and\ \citenamefont {Moore}}]{Arnold:2006fz}%
  \BibitemOpen
  \bibfield  {author} {\bibinfo {author} {\bibfnamefont {Peter~Brockway}\
  \bibnamefont {Arnold}}, \bibinfo {author} {\bibfnamefont {Caglar}\
  \bibnamefont {Dogan}}, \ and\ \bibinfo {author} {\bibfnamefont {Guy~D.}\
  \bibnamefont {Moore}},\ }\bibfield  {title} {\enquote {\bibinfo {title} {{The
  Bulk Viscosity of High-Temperature QCD}},}\ }\href {\doibase
  10.1103/PhysRevD.74.085021} {\bibfield  {journal} {\bibinfo  {journal} {Phys.
  Rev.}\ }\textbf {\bibinfo {volume} {D74}},\ \bibinfo {pages} {085021}
  (\bibinfo {year} {2006})},\ \Eprint {http://arxiv.org/abs/hep-ph/0608012}
  {arXiv:hep-ph/0608012 [hep-ph]} \BibitemShut {NoStop}%
%%CITATION = HEP-PH/0608012;%%
\bibitem [{\citenamefont {Buchel}(2008)}]{Buchel:2007mf}%
  \BibitemOpen
  \bibfield  {author} {\bibinfo {author} {\bibfnamefont {Alex}\ \bibnamefont
  {Buchel}},\ }\bibfield  {title} {\enquote {\bibinfo {title} {{Bulk viscosity
  of gauge theory plasma at strong coupling}},}\ }\href {\doibase
  10.1016/j.physletb.2008.03.069} {\bibfield  {journal} {\bibinfo  {journal}
  {Phys. Lett.}\ }\textbf {\bibinfo {volume} {B663}},\ \bibinfo {pages}
  {286--289} (\bibinfo {year} {2008})},\ \Eprint
  {http://arxiv.org/abs/0708.3459} {arXiv:0708.3459 [hep-th]} \BibitemShut
  {NoStop}%
%%CITATION = ARXIV:0708.3459;%%
\bibitem [{\citenamefont {Moore}\ and\ \citenamefont
  {Saremi}(2008)}]{Moore:2008ws}%
  \BibitemOpen
  \bibfield  {author} {\bibinfo {author} {\bibfnamefont {Guy~D.}\ \bibnamefont
  {Moore}}\ and\ \bibinfo {author} {\bibfnamefont {Omid}\ \bibnamefont
  {Saremi}},\ }\bibfield  {title} {\enquote {\bibinfo {title} {{Bulk viscosity
  and spectral functions in QCD}},}\ }\href {\doibase
  10.1088/1126-6708/2008/09/015} {\bibfield  {journal} {\bibinfo  {journal}
  {JHEP}\ }\textbf {\bibinfo {volume} {09}},\ \bibinfo {pages} {015} (\bibinfo
  {year} {2008})},\ \Eprint {http://arxiv.org/abs/0805.4201} {arXiv:0805.4201
  [hep-ph]} \BibitemShut {NoStop}%
%%CITATION = ARXIV:0805.4201;%%
\bibitem [{\citenamefont {Lu}\ and\ \citenamefont {Moore}(2011)}]{Lu:2011df}%
  \BibitemOpen
  \bibfield  {author} {\bibinfo {author} {\bibfnamefont {Egang}\ \bibnamefont
  {Lu}}\ and\ \bibinfo {author} {\bibfnamefont {Guy~D.}\ \bibnamefont
  {Moore}},\ }\bibfield  {title} {\enquote {\bibinfo {title} {{The Bulk
  Viscosity of a Pion Gas}},}\ }\href {\doibase 10.1103/PhysRevC.83.044901}
  {\bibfield  {journal} {\bibinfo  {journal} {Phys. Rev.}\ }\textbf {\bibinfo
  {volume} {C83}},\ \bibinfo {pages} {044901} (\bibinfo {year} {2011})},\
  \Eprint {http://arxiv.org/abs/1102.0017} {arXiv:1102.0017 [hep-ph]}
  \BibitemShut {NoStop}%
%%CITATION = ARXIV:1102.0017;%%
\bibitem [{\citenamefont {Dobado}\ \emph {et~al.}(2011)\citenamefont {Dobado},
  \citenamefont {Llanes-Estrada},\ and\ \citenamefont
  {Torres-Rincon}}]{Dobado:2011qu}%
  \BibitemOpen
  \bibfield  {author} {\bibinfo {author} {\bibfnamefont {Antonio}\ \bibnamefont
  {Dobado}}, \bibinfo {author} {\bibfnamefont {Felipe~J.}\ \bibnamefont
  {Llanes-Estrada}}, \ and\ \bibinfo {author} {\bibfnamefont {Juan~M.}\
  \bibnamefont {Torres-Rincon}},\ }\bibfield  {title} {\enquote {\bibinfo
  {title} {{Bulk viscosity of low-temperature strongly interacting matter}},}\
  }\href {\doibase 10.1016/j.physletb.2011.06.059} {\bibfield  {journal}
  {\bibinfo  {journal} {Phys. Lett.}\ }\textbf {\bibinfo {volume} {B702}},\
  \bibinfo {pages} {43--48} (\bibinfo {year} {2011})},\ \Eprint
  {http://arxiv.org/abs/1103.0735} {arXiv:1103.0735 [hep-ph]} \BibitemShut
  {NoStop}%
%%CITATION = ARXIV:1103.0735;%%
\bibitem [{\citenamefont {Meyer}(2011)}]{Meyer:2011gj}%
  \BibitemOpen
  \bibfield  {author} {\bibinfo {author} {\bibfnamefont {Harvey~B.}\
  \bibnamefont {Meyer}},\ }\bibfield  {title} {\enquote {\bibinfo {title}
  {{Transport Properties of the Quark-Gluon Plasma: A Lattice QCD
  Perspective}},}\ }\href {\doibase 10.1140/epja/i2011-11086-3} {\bibfield
  {journal} {\bibinfo  {journal} {Eur. Phys. J.}\ }\textbf {\bibinfo {volume}
  {A47}},\ \bibinfo {pages} {86} (\bibinfo {year} {2011})},\ \Eprint
  {http://arxiv.org/abs/1104.3708} {arXiv:1104.3708 [hep-lat]} \BibitemShut
  {NoStop}%
%%CITATION = ARXIV:1104.3708;%%
\bibitem [{\citenamefont {Czajka}\ \emph {et~al.}(2019)\citenamefont {Czajka},
  \citenamefont {Dasgupta}, \citenamefont {Gale}, \citenamefont {Jeon},
  \citenamefont {Misra}, \citenamefont {Richard},\ and\ \citenamefont
  {Sil}}]{Czajka:2018bod}%
  \BibitemOpen
  \bibfield  {author} {\bibinfo {author} {\bibfnamefont {Alina}\ \bibnamefont
  {Czajka}}, \bibinfo {author} {\bibfnamefont {Keshav}\ \bibnamefont
  {Dasgupta}}, \bibinfo {author} {\bibfnamefont {Charles}\ \bibnamefont
  {Gale}}, \bibinfo {author} {\bibfnamefont {Sangyong}\ \bibnamefont {Jeon}},
  \bibinfo {author} {\bibfnamefont {Aalok}\ \bibnamefont {Misra}}, \bibinfo
  {author} {\bibfnamefont {Michael}\ \bibnamefont {Richard}}, \ and\ \bibinfo
  {author} {\bibfnamefont {Karunava}\ \bibnamefont {Sil}},\ }\bibfield  {title}
  {\enquote {\bibinfo {title} {{Bulk Viscosity at Extreme Limits: From Kinetic
  Theory to Strings}},}\ }\href {\doibase 10.1007/JHEP07(2019)145} {\bibfield
  {journal} {\bibinfo  {journal} {JHEP}\ }\textbf {\bibinfo {volume} {07}},\
  \bibinfo {pages} {145} (\bibinfo {year} {2019})},\ \Eprint
  {http://arxiv.org/abs/1807.04713} {arXiv:1807.04713 [hep-th]} \BibitemShut
  {NoStop}%
%%CITATION = ARXIV:1807.04713;%%
\bibitem [{\citenamefont {Torrieri}\ \emph {et~al.}(2008)\citenamefont
  {Torrieri}, \citenamefont {Tomasik},\ and\ \citenamefont
  {Mishustin}}]{Torrieri:2007fb}%
  \BibitemOpen
  \bibfield  {author} {\bibinfo {author} {\bibfnamefont {Giorgio}\ \bibnamefont
  {Torrieri}}, \bibinfo {author} {\bibfnamefont {Boris}\ \bibnamefont
  {Tomasik}}, \ and\ \bibinfo {author} {\bibfnamefont {Igor}\ \bibnamefont
  {Mishustin}},\ }\bibfield  {title} {\enquote {\bibinfo {title} {{Bulk
  Viscosity driven clusterization of quark-gluon plasma and early freeze-out in
  relativistic heavy-ion collisions}},}\ }\href {\doibase
  10.1103/PhysRevC.77.034903} {\bibfield  {journal} {\bibinfo  {journal} {Phys.
  Rev.}\ }\textbf {\bibinfo {volume} {C77}},\ \bibinfo {pages} {034903}
  (\bibinfo {year} {2008})},\ \Eprint {http://arxiv.org/abs/0707.4405}
  {arXiv:0707.4405 [nucl-th]} \BibitemShut {NoStop}%
%%CITATION = ARXIV:0707.4405;%%
\bibitem [{\citenamefont {Karsch}\ \emph {et~al.}(2008)\citenamefont {Karsch},
  \citenamefont {Kharzeev},\ and\ \citenamefont {Tuchin}}]{Karsch:2007jc}%
  \BibitemOpen
  \bibfield  {author} {\bibinfo {author} {\bibfnamefont {Frithjof}\
  \bibnamefont {Karsch}}, \bibinfo {author} {\bibfnamefont {Dmitri}\
  \bibnamefont {Kharzeev}}, \ and\ \bibinfo {author} {\bibfnamefont {Kirill}\
  \bibnamefont {Tuchin}},\ }\bibfield  {title} {\enquote {\bibinfo {title}
  {{Universal properties of bulk viscosity near the QCD phase transition}},}\
  }\href {\doibase 10.1016/j.physletb.2008.01.080} {\bibfield  {journal}
  {\bibinfo  {journal} {Phys. Lett.}\ }\textbf {\bibinfo {volume} {B663}},\
  \bibinfo {pages} {217--221} (\bibinfo {year} {2008})},\ \Eprint
  {http://arxiv.org/abs/0711.0914} {arXiv:0711.0914 [hep-ph]} \BibitemShut
  {NoStop}%
%%CITATION = ARXIV:0711.0914;%%
\bibitem [{\citenamefont {Romatschke}\ and\ \citenamefont
  {Son}(2009)}]{Romatschke:2009ng}%
  \BibitemOpen
  \bibfield  {author} {\bibinfo {author} {\bibfnamefont {Paul}\ \bibnamefont
  {Romatschke}}\ and\ \bibinfo {author} {\bibfnamefont {Dam~Thanh}\
  \bibnamefont {Son}},\ }\bibfield  {title} {\enquote {\bibinfo {title}
  {{Spectral sum rules for the quark-gluon plasma}},}\ }\href {\doibase
  10.1103/PhysRevD.80.065021} {\bibfield  {journal} {\bibinfo  {journal} {Phys.
  Rev.}\ }\textbf {\bibinfo {volume} {D80}},\ \bibinfo {pages} {065021}
  (\bibinfo {year} {2009})},\ \Eprint {http://arxiv.org/abs/0903.3946}
  {arXiv:0903.3946 [hep-ph]} \BibitemShut {NoStop}%
%%CITATION = ARXIV:0903.3946;%%
\bibitem [{\citenamefont {Moreland}\ \emph {et~al.}(2018)\citenamefont
  {Moreland}, \citenamefont {Bernhard},\ and\ \citenamefont
  {Bass}}]{Moreland:2018gsh}%
  \BibitemOpen
  \bibfield  {author} {\bibinfo {author} {\bibfnamefont {J.~Scott}\
  \bibnamefont {Moreland}}, \bibinfo {author} {\bibfnamefont {Jonah~E.}\
  \bibnamefont {Bernhard}}, \ and\ \bibinfo {author} {\bibfnamefont
  {Steffen~A.}\ \bibnamefont {Bass}},\ }\bibfield  {title} {\enquote {\bibinfo
  {title} {{Estimating initial state and quark-gluon plasma medium properties
  using a hybrid model with nucleon substructure calibrated to $p$-Pb and Pb-Pb
  collisions at $\sqrt{s_\mathrm{NN}}=5.02$ TeV}},}\ }\href@noop {} {\
  (\bibinfo {year} {2018})},\ \Eprint {http://arxiv.org/abs/1808.02106}
  {arXiv:1808.02106 [nucl-th]} \BibitemShut {NoStop}%
%%CITATION = ARXIV:1808.02106;%%
\bibitem [{\citenamefont {Hagedorn}(1965)}]{Hagedorn:1965st}%
  \BibitemOpen
  \bibfield  {author} {\bibinfo {author} {\bibfnamefont {R.}~\bibnamefont
  {Hagedorn}},\ }\bibfield  {title} {\enquote {\bibinfo {title} {{Statistical
  thermodynamics of strong interactions at high-energies}},}\ }\href@noop {}
  {\bibfield  {journal} {\bibinfo  {journal} {Nuovo Cim. Suppl.}\ }\textbf
  {\bibinfo {volume} {3}},\ \bibinfo {pages} {147--186} (\bibinfo {year}
  {1965})}\BibitemShut {NoStop}%
%%CITATION = NUCUA,3,147;%%
\bibitem [{\citenamefont {Noronha-Hostler}\ \emph {et~al.}(2009)\citenamefont
  {Noronha-Hostler}, \citenamefont {Noronha},\ and\ \citenamefont
  {Greiner}}]{NoronhaHostler:2008ju}%
  \BibitemOpen
  \bibfield  {author} {\bibinfo {author} {\bibfnamefont {Jacquelyn}\
  \bibnamefont {Noronha-Hostler}}, \bibinfo {author} {\bibfnamefont {Jorge}\
  \bibnamefont {Noronha}}, \ and\ \bibinfo {author} {\bibfnamefont {Carsten}\
  \bibnamefont {Greiner}},\ }\bibfield  {title} {\enquote {\bibinfo {title}
  {{Transport Coefficients of Hadronic Matter near T(c)}},}\ }\href {\doibase
  10.1103/PhysRevLett.103.172302} {\bibfield  {journal} {\bibinfo  {journal}
  {Phys. Rev. Lett.}\ }\textbf {\bibinfo {volume} {103}},\ \bibinfo {pages}
  {172302} (\bibinfo {year} {2009})},\ \Eprint {http://arxiv.org/abs/0811.1571}
  {arXiv:0811.1571 [nucl-th]} \BibitemShut {NoStop}%
%%CITATION = ARXIV:0811.1571;%%
\bibitem [{\citenamefont {Ryu}\ \emph {et~al.}(2018)\citenamefont {Ryu},
  \citenamefont {Paquet}, \citenamefont {Shen}, \citenamefont {Denicol},
  \citenamefont {Schenke}, \citenamefont {Jeon},\ and\ \citenamefont
  {Gale}}]{PhysRevC.97.034910}%
  \BibitemOpen
  \bibfield  {author} {\bibinfo {author} {\bibfnamefont {Sangwook}\
  \bibnamefont {Ryu}}, \bibinfo {author} {\bibfnamefont {Jean-Fran\ifmmode
  \mbox{\c{c}}\else~\c{c}\fi{}ois}\ \bibnamefont {Paquet}}, \bibinfo {author}
  {\bibfnamefont {Chun}\ \bibnamefont {Shen}}, \bibinfo {author} {\bibfnamefont
  {Gabriel}\ \bibnamefont {Denicol}}, \bibinfo {author} {\bibfnamefont
  {Bj$\ddot{\rm o}$rn}\ \bibnamefont {Schenke}}, \bibinfo {author}
  {\bibfnamefont {Sangyong}\ \bibnamefont {Jeon}}, \ and\ \bibinfo {author}
  {\bibfnamefont {Charles}\ \bibnamefont {Gale}},\ }\bibfield  {title}
  {\enquote {\bibinfo {title} {Effects of bulk viscosity and hadronic
  rescattering in heavy ion collisions at energies available at the bnl
  relativistic heavy ion collider and at the cern large hadron collider},}\
  }\href {\doibase 10.1103/PhysRevC.97.034910} {\bibfield  {journal} {\bibinfo
  {journal} {Phys. Rev. C}\ }\textbf {\bibinfo {volume} {97}},\ \bibinfo
  {pages} {034910} (\bibinfo {year} {2018})}\BibitemShut {NoStop}%
\bibitem [{\citenamefont {Schenke}\ \emph
  {et~al.}(2019{\natexlab{a}})\citenamefont {Schenke}, \citenamefont {Shen},\
  and\ \citenamefont {Tribedy}}]{Schenke:2019ruo}%
  \BibitemOpen
  \bibfield  {author} {\bibinfo {author} {\bibfnamefont {Bj$\ddot{\rm o}$rn}\
  \bibnamefont {Schenke}}, \bibinfo {author} {\bibfnamefont {Chun}\
  \bibnamefont {Shen}}, \ and\ \bibinfo {author} {\bibfnamefont {Prithwish}\
  \bibnamefont {Tribedy}},\ }\bibfield  {title} {\enquote {\bibinfo {title}
  {{Multi-particle and charge-dependent azimuthal correlations in heavy-ion
  collisions at the Relativistic Heavy-Ion Collider}},}\ }\href {\doibase
  10.1103/PhysRevC.99.044908} {\bibfield  {journal} {\bibinfo  {journal} {Phys.
  Rev.}\ }\textbf {\bibinfo {volume} {C99}},\ \bibinfo {pages} {044908}
  (\bibinfo {year} {2019}{\natexlab{a}})},\ \Eprint
  {http://arxiv.org/abs/1901.04378} {arXiv:1901.04378 [nucl-th]} \BibitemShut
  {NoStop}%
%%CITATION = ARXIV:1901.04378;%%
\bibitem [{\citenamefont {Schenke}\ \emph {et~al.}(2012)\citenamefont
  {Schenke}, \citenamefont {Tribedy},\ and\ \citenamefont
  {Venugopalan}}]{Schenke:2012wb}%
  \BibitemOpen
  \bibfield  {author} {\bibinfo {author} {\bibfnamefont {Bj$\ddot{\rm o}$rn}\
  \bibnamefont {Schenke}}, \bibinfo {author} {\bibfnamefont {Prithwish}\
  \bibnamefont {Tribedy}}, \ and\ \bibinfo {author} {\bibfnamefont {Raju}\
  \bibnamefont {Venugopalan}},\ }\bibfield  {title} {\enquote {\bibinfo {title}
  {{Fluctuating Glasma initial conditions and flow in heavy ion collisions}},}\
  }\href {\doibase 10.1103/PhysRevLett.108.252301} {\bibfield  {journal}
  {\bibinfo  {journal} {Phys. Rev. Lett.}\ }\textbf {\bibinfo {volume} {108}},\
  \bibinfo {pages} {252301} (\bibinfo {year} {2012})},\ \Eprint
  {http://arxiv.org/abs/1202.6646} {arXiv:1202.6646 [nucl-th]} \BibitemShut
  {NoStop}%
\bibitem [{\citenamefont {Schenke}\ \emph
  {et~al.}(2019{\natexlab{b}})\citenamefont {Schenke}, \citenamefont {Shen},\
  and\ \citenamefont {Tribedy}}]{Schenke:2018fci}%
  \BibitemOpen
  \bibfield  {author} {\bibinfo {author} {\bibfnamefont {Bj$\ddot{\rm o}$rn}\
  \bibnamefont {Schenke}}, \bibinfo {author} {\bibfnamefont {Chun}\
  \bibnamefont {Shen}}, \ and\ \bibinfo {author} {\bibfnamefont {Prithwish}\
  \bibnamefont {Tribedy}},\ }\bibfield  {title} {\enquote {\bibinfo {title}
  {{Features of the IP-Glasma}},}\ }\bibfield  {booktitle} {\emph {\bibinfo
  {booktitle} {{Proceedings, 27th International Conference on Ultrarelativistic
  Nucleus-Nucleus Collisions (Quark Matter 2018): Venice, Italy, May 14-19,
  2018}}},\ }\href {\doibase 10.1016/j.nuclphysa.2018.08.015} {\bibfield
  {journal} {\bibinfo  {journal} {Nucl. Phys.}\ }\textbf {\bibinfo {volume}
  {A982}},\ \bibinfo {pages} {435--438} (\bibinfo {year}
  {2019}{\natexlab{b}})},\ \Eprint {http://arxiv.org/abs/1807.05205}
  {arXiv:1807.05205 [nucl-th]} \BibitemShut {NoStop}%
%%CITATION = ARXIV:1807.05205;%%
\bibitem [{\citenamefont {Loizides}\ and\ \citenamefont
  {Morsch}(2017)}]{Morsch:2017brb}%
  \BibitemOpen
  \bibfield  {author} {\bibinfo {author} {\bibfnamefont {Constantin}\
  \bibnamefont {Loizides}}\ and\ \bibinfo {author} {\bibfnamefont {Andreas}\
  \bibnamefont {Morsch}},\ }\bibfield  {title} {\enquote {\bibinfo {title}
  {{Absence of jet quenching in peripheral nucleus–nucleus collisions}},}\
  }\href {\doibase 10.1016/j.physletb.2017.09.002} {\bibfield  {journal}
  {\bibinfo  {journal} {Phys. Lett.}\ }\textbf {\bibinfo {volume} {B773}},\
  \bibinfo {pages} {408--411} (\bibinfo {year} {2017})},\ \Eprint
  {http://arxiv.org/abs/1705.08856} {arXiv:1705.08856 [nucl-ex]} \BibitemShut
  {NoStop}%
%%CITATION = ARXIV:1705.08856;%%
\bibitem [{\citenamefont {Acharya}\ \emph {et~al.}(2019)\citenamefont {Acharya}
  \emph {et~al.}}]{Acharya:2018njl}%
  \BibitemOpen
  \bibfield  {author} {\bibinfo {author} {\bibfnamefont {Shreyasi}\
  \bibnamefont {Acharya}} \emph {et~al.} (\bibinfo {collaboration} {ALICE}),\
  }\bibfield  {title} {\enquote {\bibinfo {title} {{Analysis of the apparent
  nuclear modification in peripheral Pb–Pb collisions at 5.02 TeV}},}\ }\href
  {\doibase 10.1016/j.physletb.2019.04.047} {\bibfield  {journal} {\bibinfo
  {journal} {Phys. Lett.}\ }\textbf {\bibinfo {volume} {B793}},\ \bibinfo
  {pages} {420--432} (\bibinfo {year} {2019})},\ \Eprint
  {http://arxiv.org/abs/1805.05212} {arXiv:1805.05212 [nucl-ex]} \BibitemShut
  {NoStop}%
%%CITATION = ARXIV:1805.05212;%%
\bibitem [{\citenamefont {Schenke}\ \emph
  {et~al.}(2019{\natexlab{c}})\citenamefont {Schenke}, \citenamefont {Shen},\
  and\ \citenamefont {Tribedy}}]{Schenke:2019pmk}%
  \BibitemOpen
  \bibfield  {author} {\bibinfo {author} {\bibfnamefont {Bj$\ddot{\rm o}$rn}\
  \bibnamefont {Schenke}}, \bibinfo {author} {\bibfnamefont {Chun}\
  \bibnamefont {Shen}}, \ and\ \bibinfo {author} {\bibfnamefont {Prithwish}\
  \bibnamefont {Tribedy}},\ }\bibfield  {title} {\enquote {\bibinfo {title}
  {{Hybrid Color Glass Condensate and hydrodynamic description of the
  Relativistic Heavy Ion Collider small system scan}},}\ }\href@noop {} {\
  (\bibinfo {year} {2019}{\natexlab{c}})},\ \Eprint
  {http://arxiv.org/abs/1908.06212} {arXiv:1908.06212 [nucl-th]} \BibitemShut
  {NoStop}%
%%CITATION = ARXIV:1908.06212;%%
\bibitem [{\citenamefont {Denicol}\ \emph {et~al.}(2015)\citenamefont
  {Denicol}, \citenamefont {Gale},\ and\ \citenamefont
  {Jeon}}]{Denicol:2015bpa}%
  \BibitemOpen
  \bibfield  {author} {\bibinfo {author} {\bibfnamefont {Gabriel~S.}\
  \bibnamefont {Denicol}}, \bibinfo {author} {\bibfnamefont {Charles}\
  \bibnamefont {Gale}}, \ and\ \bibinfo {author} {\bibfnamefont {Sangyong}\
  \bibnamefont {Jeon}},\ }\bibfield  {title} {\enquote {\bibinfo {title} {{The
  domain of validity of fluid dynamics and the onset of cavitation in
  ultrarelativistic heavy ion collisions}},}\ }\bibfield  {booktitle} {\emph
  {\bibinfo {booktitle} {{Proceedings, 9th International Workshop on Critical
  Point and Onset of Deconfinement (CPOD 2014): Bielefeld, Germany, November
  17-21, 2014}}},\ }\href {\doibase 10.22323/1.217.0033} {\bibfield  {journal}
  {\bibinfo  {journal} {PoS}\ }\textbf {\bibinfo {volume} {CPOD2014}},\
  \bibinfo {pages} {033} (\bibinfo {year} {2015})},\ \Eprint
  {http://arxiv.org/abs/1503.00531} {arXiv:1503.00531 [nucl-th]} \BibitemShut
  {NoStop}%
%%CITATION = ARXIV:1503.00531;%%
\bibitem [{\citenamefont {Cooper}\ and\ \citenamefont
  {Frye}(1974)}]{Cooper:1974mv}%
  \BibitemOpen
  \bibfield  {author} {\bibinfo {author} {\bibfnamefont {Fred}\ \bibnamefont
  {Cooper}}\ and\ \bibinfo {author} {\bibfnamefont {Graham}\ \bibnamefont
  {Frye}},\ }\bibfield  {title} {\enquote {\bibinfo {title} {{Comment on the
  Single Particle Distribution in the Hydrodynamic and Statistical
  Thermodynamic Models of Multiparticle Production}},}\ }\href {\doibase
  10.1103/PhysRevD.10.186} {\bibfield  {journal} {\bibinfo  {journal} {Phys.
  Rev.}\ }\textbf {\bibinfo {volume} {D10}},\ \bibinfo {pages} {186} (\bibinfo
  {year} {1974})}\BibitemShut {NoStop}%
%%CITATION = PHRVA,D10,186;%%
\bibitem [{\citenamefont {Teaney}(2003)}]{Teaney:2003kp}%
  \BibitemOpen
  \bibfield  {author} {\bibinfo {author} {\bibfnamefont {Derek}\ \bibnamefont
  {Teaney}},\ }\bibfield  {title} {\enquote {\bibinfo {title} {{The Effects of
  viscosity on spectra, elliptic flow, and HBT radii}},}\ }\href {\doibase
  10.1103/PhysRevC.68.034913} {\bibfield  {journal} {\bibinfo  {journal} {Phys.
  Rev.}\ }\textbf {\bibinfo {volume} {C68}},\ \bibinfo {pages} {034913}
  (\bibinfo {year} {2003})},\ \Eprint {http://arxiv.org/abs/nucl-th/0301099}
  {arXiv:nucl-th/0301099 [nucl-th]} \BibitemShut {NoStop}%
%%CITATION = NUCL-TH/0301099;%%
\bibitem [{\citenamefont {Strickland}(2014)}]{Strickland:2014pga}%
  \BibitemOpen
  \bibfield  {author} {\bibinfo {author} {\bibfnamefont {Michael}\ \bibnamefont
  {Strickland}},\ }\bibfield  {title} {\enquote {\bibinfo {title} {{Anisotropic
  Hydrodynamics: Three lectures}},}\ }\bibfield  {booktitle} {\emph {\bibinfo
  {booktitle} {{54th Cracow School of Theoretical Physics: QCD meets
  experiment: Zakopane, Poland, June 12-20, 2014}}},\ }\href {\doibase
  10.5506/APhysPolB.45.2355} {\bibfield  {journal} {\bibinfo  {journal} {Acta
  Phys. Polon.}\ }\textbf {\bibinfo {volume} {B45}},\ \bibinfo {pages}
  {2355--2394} (\bibinfo {year} {2014})},\ \Eprint
  {http://arxiv.org/abs/1410.5786} {arXiv:1410.5786 [nucl-th]} \BibitemShut
  {NoStop}%
%%CITATION = ARXIV:1410.5786;%%
\bibitem [{\citenamefont {Molnar}\ and\ \citenamefont
  {Wolff}(2017)}]{Molnar:2014fva}%
  \BibitemOpen
  \bibfield  {author} {\bibinfo {author} {\bibfnamefont {Denes}\ \bibnamefont
  {Molnar}}\ and\ \bibinfo {author} {\bibfnamefont {Zack}\ \bibnamefont
  {Wolff}},\ }\bibfield  {title} {\enquote {\bibinfo {title} {{Self-consistent
  conversion of a viscous fluid to particles}},}\ }\href {\doibase
  10.1103/PhysRevC.95.024903} {\bibfield  {journal} {\bibinfo  {journal} {Phys.
  Rev.}\ }\textbf {\bibinfo {volume} {C95}},\ \bibinfo {pages} {024903}
  (\bibinfo {year} {2017})},\ \Eprint {http://arxiv.org/abs/1404.7850}
  {arXiv:1404.7850 [nucl-th]} \BibitemShut {NoStop}%
%%CITATION = ARXIV:1404.7850;%%
\bibitem [{\citenamefont {Bernhard}\ \emph {et~al.}(2016)\citenamefont
  {Bernhard}, \citenamefont {Moreland}, \citenamefont {Bass}, \citenamefont
  {Liu},\ and\ \citenamefont {Heinz}}]{Bernhard:2016tnd}%
  \BibitemOpen
  \bibfield  {author} {\bibinfo {author} {\bibfnamefont {Jonah~E.}\
  \bibnamefont {Bernhard}}, \bibinfo {author} {\bibfnamefont {J.~Scott}\
  \bibnamefont {Moreland}}, \bibinfo {author} {\bibfnamefont {Steffen~A.}\
  \bibnamefont {Bass}}, \bibinfo {author} {\bibfnamefont {Jia}\ \bibnamefont
  {Liu}}, \ and\ \bibinfo {author} {\bibfnamefont {Ulrich}\ \bibnamefont
  {Heinz}},\ }\bibfield  {title} {\enquote {\bibinfo {title} {{Applying
  Bayesian parameter estimation to relativistic heavy-ion collisions:
  simultaneous characterization of the initial state and quark-gluon plasma
  medium}},}\ }\href {\doibase 10.1103/PhysRevC.94.024907} {\bibfield
  {journal} {\bibinfo  {journal} {Phys. Rev.}\ }\textbf {\bibinfo {volume}
  {C94}},\ \bibinfo {pages} {024907} (\bibinfo {year} {2016})},\ \Eprint
  {http://arxiv.org/abs/1605.03954} {arXiv:1605.03954 [nucl-th]} \BibitemShut
  {NoStop}%
%%CITATION = ARXIV:1605.03954;%%
\bibitem [{\citenamefont {Weller}\ and\ \citenamefont
  {Romatschke}(2017)}]{Weller:2017tsr}%
  \BibitemOpen
  \bibfield  {author} {\bibinfo {author} {\bibfnamefont {Ryan~D.}\ \bibnamefont
  {Weller}}\ and\ \bibinfo {author} {\bibfnamefont {Paul}\ \bibnamefont
  {Romatschke}},\ }\bibfield  {title} {\enquote {\bibinfo {title} {{One fluid
  to rule them all: viscous hydrodynamic description of event-by-event central
  p+p, p+Pb and Pb+Pb collisions at $\sqrt{s}=5.02$ TeV}},}\ }\href {\doibase
  10.1016/j.physletb.2017.09.077} {\bibfield  {journal} {\bibinfo  {journal}
  {Phys. Lett.}\ }\textbf {\bibinfo {volume} {B774}},\ \bibinfo {pages}
  {351--356} (\bibinfo {year} {2017})},\ \Eprint
  {http://arxiv.org/abs/1701.07145} {arXiv:1701.07145 [nucl-th]} \BibitemShut
  {NoStop}%
%%CITATION = ARXIV:1701.07145;%%
\bibitem [{\citenamefont {van~der Schee}\ \emph {et~al.}(2013)\citenamefont
  {van~der Schee}, \citenamefont {Romatschke},\ and\ \citenamefont
  {Pratt}}]{vanderSchee:2013pia}%
  \BibitemOpen
  \bibfield  {author} {\bibinfo {author} {\bibfnamefont {Wilke}\ \bibnamefont
  {van~der Schee}}, \bibinfo {author} {\bibfnamefont {Paul}\ \bibnamefont
  {Romatschke}}, \ and\ \bibinfo {author} {\bibfnamefont {Scott}\ \bibnamefont
  {Pratt}},\ }\bibfield  {title} {\enquote {\bibinfo {title} {{Fully Dynamical
  Simulation of Central Nuclear Collisions}},}\ }\href {\doibase
  10.1103/PhysRevLett.111.222302} {\bibfield  {journal} {\bibinfo  {journal}
  {Phys. Rev. Lett.}\ }\textbf {\bibinfo {volume} {111}},\ \bibinfo {pages}
  {222302} (\bibinfo {year} {2013})},\ \Eprint {http://arxiv.org/abs/1307.2539}
  {arXiv:1307.2539 [nucl-th]} \BibitemShut {NoStop}%
%%CITATION = ARXIV:1307.2539;%%
\bibitem [{\citenamefont {Rajagopal}\ and\ \citenamefont
  {Tripuraneni}(2010)}]{Rajagopal:2009yw}%
  \BibitemOpen
  \bibfield  {author} {\bibinfo {author} {\bibfnamefont {Krishna}\ \bibnamefont
  {Rajagopal}}\ and\ \bibinfo {author} {\bibfnamefont {Nilesh}\ \bibnamefont
  {Tripuraneni}},\ }\bibfield  {title} {\enquote {\bibinfo {title} {{Bulk
  Viscosity and Cavitation in Boost-Invariant Hydrodynamic Expansion}},}\
  }\href {\doibase 10.1007/JHEP03(2010)018} {\bibfield  {journal} {\bibinfo
  {journal} {JHEP}\ }\textbf {\bibinfo {volume} {03}},\ \bibinfo {pages} {018}
  (\bibinfo {year} {2010})},\ \Eprint {http://arxiv.org/abs/0908.1785}
  {arXiv:0908.1785 [hep-ph]} \BibitemShut {NoStop}%
%%CITATION = ARXIV:0908.1785;%%
\bibitem [{\citenamefont {Habich}\ \emph {et~al.}(2016)\citenamefont {Habich},
  \citenamefont {Miller}, \citenamefont {Romatschke},\ and\ \citenamefont
  {Xiang}}]{Habich:2015rtj}%
  \BibitemOpen
  \bibfield  {author} {\bibinfo {author} {\bibfnamefont {M.}~\bibnamefont
  {Habich}}, \bibinfo {author} {\bibfnamefont {G.~A.}\ \bibnamefont {Miller}},
  \bibinfo {author} {\bibfnamefont {P.}~\bibnamefont {Romatschke}}, \ and\
  \bibinfo {author} {\bibfnamefont {W.}~\bibnamefont {Xiang}},\ }\bibfield
  {title} {\enquote {\bibinfo {title} {{Testing hydrodynamic descriptions of
  p+p collisions at $\sqrt{s}=7$ TeV}},}\ }\href {\doibase
  10.1140/epjc/s10052-016-4237-z} {\bibfield  {journal} {\bibinfo  {journal}
  {Eur. Phys. J.}\ }\textbf {\bibinfo {volume} {C76}},\ \bibinfo {pages} {408}
  (\bibinfo {year} {2016})},\ \Eprint {http://arxiv.org/abs/1512.05354}
  {arXiv:1512.05354 [nucl-th]} \BibitemShut {NoStop}%
%%CITATION = ARXIV:1512.05354;%%
\bibitem [{\citenamefont {Habich}\ and\ \citenamefont
  {Romatschke}(2014)}]{Habich:2014tpa}%
  \BibitemOpen
  \bibfield  {author} {\bibinfo {author} {\bibfnamefont {Mathis}\ \bibnamefont
  {Habich}}\ and\ \bibinfo {author} {\bibfnamefont {Paul}\ \bibnamefont
  {Romatschke}},\ }\bibfield  {title} {\enquote {\bibinfo {title} {{Onset of
  cavitation in the quark-gluon plasma}},}\ }\href {\doibase
  10.1007/JHEP12(2014)054} {\bibfield  {journal} {\bibinfo  {journal} {JHEP}\
  }\textbf {\bibinfo {volume} {12}},\ \bibinfo {pages} {054} (\bibinfo {year}
  {2014})},\ \Eprint {http://arxiv.org/abs/1405.1978} {arXiv:1405.1978
  [hep-ph]} \BibitemShut {NoStop}%
%%CITATION = ARXIV:1405.1978;%%
\bibitem [{\citenamefont {Schenke}\ \emph {et~al.}(2010)\citenamefont
  {Schenke}, \citenamefont {Jeon},\ and\ \citenamefont
  {Gale}}]{Schenke:2010nt}%
  \BibitemOpen
  \bibfield  {author} {\bibinfo {author} {\bibfnamefont {Bj$\ddot{\rm o}$rn}\
  \bibnamefont {Schenke}}, \bibinfo {author} {\bibfnamefont {Sangyong}\
  \bibnamefont {Jeon}}, \ and\ \bibinfo {author} {\bibfnamefont {Charles}\
  \bibnamefont {Gale}},\ }\bibfield  {title} {\enquote {\bibinfo {title}
  {{(3+1)D hydrodynamic simulation of relativistic heavy-ion collisions}},}\
  }\href {\doibase 10.1103/PhysRevC.82.014903} {\bibfield  {journal} {\bibinfo
  {journal} {Phys. Rev.}\ }\textbf {\bibinfo {volume} {C82}},\ \bibinfo {pages}
  {014903} (\bibinfo {year} {2010})},\ \Eprint {http://arxiv.org/abs/1004.1408}
  {arXiv:1004.1408 [hep-ph]} \BibitemShut {NoStop}%
%%CITATION = ARXIV:1004.1408;%%
\bibitem [{\citenamefont {Habich}\ \emph {et~al.}(2015)\citenamefont {Habich},
  \citenamefont {Nagle},\ and\ \citenamefont {Romatschke}}]{Habich:2014jna}%
  \BibitemOpen
  \bibfield  {author} {\bibinfo {author} {\bibfnamefont {M.}~\bibnamefont
  {Habich}}, \bibinfo {author} {\bibfnamefont {J.~L.}\ \bibnamefont {Nagle}}, \
  and\ \bibinfo {author} {\bibfnamefont {P.}~\bibnamefont {Romatschke}},\
  }\bibfield  {title} {\enquote {\bibinfo {title} {{Particle spectra and HBT
  radii for simulated central nuclear collisions of C$+$C, Al$+$Al, Cu$+$Cu,
  Au$+$Au, and Pb$+$Pb from $\sqrt{s}=62.4$--$2760$ GeV}},}\ }\href {\doibase
  10.1140/epjc/s10052-014-3206-7} {\bibfield  {journal} {\bibinfo  {journal}
  {Eur. Phys. J.}\ }\textbf {\bibinfo {volume} {C75}},\ \bibinfo {pages} {15}
  (\bibinfo {year} {2015})},\ \Eprint {http://arxiv.org/abs/1409.0040}
  {arXiv:1409.0040 [nucl-th]} \BibitemShut {NoStop}%
\end{thebibliography}%

\end{document}